# A ~60 Myr periodicity is common to marine-$^{87}$Sr/$^{86}$Sr, fossil biodiversity, and large-scale sedimentation: what does the periodicity reflect?

Adrian L. Melott[1*], Richard K. Bambach[2], Kenni D. Petersen[3], and John M. McArthur[4]

*1 Department of Physics and Astronomy, University of Kansas, Lawrence, Kansas 66045 USA.*
*2 Department of Paleobiology, National Museum of Natural History, Smithsonian Institution, PO Box 37012, MRC 121, Washington, DC 20013-7012 USA.*

*3 Department of Earth Science, Aarhus University, DK-8000 Aarhus C, Denmark.*
*4. Dept. of Earth Sciences, University College London, Gower Street, London WC1E 6BT, UK*
*\* Author for correspondence, email: melott@ku.edu*

**ABSTRACT**

We find that the marine $^{87}$Sr/$^{86}$Sr record shows a significant periodicity of 59.3 ± 3 Myr. The $^{87}$Sr/$^{86}$Sr record is 171° ± 12° out of phase with a 62 (± 3) Myr periodicity previously reported in the record of marine-animal diversity. These periodicities are close to 58 (± 4) Myr cycles found for the number of gap-bounded sedimentary carbonate packages of North America We propose that these periodicities reflect the operation of a periodic "pulse of the Earth" in large-scale, Earth processes. These may be linked to mantle or plate-tectonic events, possibly uplift, which affects Earth's climate and oceans, and so the geochemistry, sedimentation, and biodiversity of the marine realm.
**Online enhancement:** Appendix on methodology and related results

**Introduction**
   It has frequently been suggested that large-scale geological phenomena may be the driver for large-scale evolutionary change that is revealed in the fossil record; for example, that the emplacement of large igneous-provinces may drive mass extinctions (e.g. Courtillot et al., 1996). Possible connections between large-scale tectonic phenomena, the fossil record, and the record of marine $^{87}Sr/^{86}Sr$ (a proxy for continental weathering) have been noted before, by casual allusion (e.g. McArthur et al. 2001), minimal discussion (Prokopf 2009) discussion that lacked quantification (Halverson et al., 2007; Purdy, 2008). Here we provide a time-series analysis of an independent signal of the geological record in an attempt to reveal more about large-scale geological events as potential drivers of biodiversity change.

   We note that a 62 ± 3 Myr periodicity exists in the record of marine animal diversity (Rhode and Muller, 2005; Cornette, 2007; Lieberman and Melott, 2007; Melott, 2008; Melott and Bambach, 2011a). This periodicity is present in three independent sets of data that we have examined that document Phanerozoic marine diversity (details in Melott and Bambach, 2011a and references therein). A periodicity of 58 ± 4 Myr is present in the temporal and area-weighted number of gap-bound sedimentary carbonate in the United States (Melott and Bambach 2011b, using data in Peters 2008). After Melott and Bambach (2011b) was accepted for publication, Meyers and Peters (2011) was published, showing a 56 ± 3 Myr periodicity in the total sediment area coverage of an expanded set of North American data. Both periodicities are significant and in phase (either quantify the closeness, or delete the adjective) with the biodiversity periodicity (Melott and Bambach, 2011b). Error bars (±) on periods given in this work signify the bandwidth of the spectral peak, i.e. an estimate of one standard deviation in the frequency distribution of power as observed on a spectral plot.

   Other reported periodicities include a ~60 Myr periodicity in the timing of emplacement of continental large igneous provinces over the last 320 Myr (Prokoph et al. 2004), which we discuss later. Almost as an aside, a ~60 Myr periodicity was noted in the record of marine $^{87}Sr/^{86}Sr$ by Prokopf et al. (2008). Neither observation was supported by significance levels of the detection, phase angle, nor extended discussion.

**Methods**
We examine periodicity using the amplitude peaks of the Fourier coefficients of time-series, a widely used method (Press et al., 2007; Muller and MacDonald, 2002). Virtually any mathematical function can be decomposed into a sum of sinusoids, but a significant indication of periodicity exists when the amplitude of some frequency stands out significantly above the others.
       Developing the arguments in Melott and Bambach (2011a) and (2011b), here we tested for periodicity in marine-$^{87}Sr/^{86}Sr$ using the LOWESS fit of McArthur and Howarth (2004) for the Phanerozoic, and the number of short-lived genera (those which endure less than 45 Myr)

given by Rohde and Muller (2005), which are reduced from the database of Sepkoski (2002). Both sets of data use are adjusted to the timescale of Gradstein et al. (2004).

The LOWESS fit of McArthur and Howarth (2004) is based contain 3875 pairs of published data, unevenly spread through Phanerozoic time. The uncertainty of positioning of peaks and troughs in the Sr curve within the timescale used (Gradstein et al. 2004) matches the quality of age given by the timescale used, and is typically ± 2 Myr for the Palaeozoic and Mesozoic, and typically < 0.3 Myr for the Cenozoic < 1 Myr. Because the residence time of Sr in the world's oceans is today around 2 Myr, marine-$^{87}$Sr/$^{86}$Sr is buffered against rapid change (i.e. timescales < 1 Myr; (Richter and Turekian 1993), it is certain that no major excursions remain unidentified.

In line with standard methods, to facilitate our Fourier analysis, we use a LOWESS curve that was detrended by fitting a cubic regression, calculating residuals about the fitted line, and normalized by dividing each value of $^{87}$Sr/$^{86}$Sr by the relevant standard deviation (see Online Appendix for details). Detrending is standard in time-series analysis, and has no effect on inherent periodicity in time-series signals other than at the period of the fit. Normalization (as defined earlier) is a mathematical convenience that permits better comparison between different sets of time-series data: To compare periodicity and phase between time-series results from different data sets, we use cross-spectral analysis. The $^{87}$Sr/$^{86}$Sr curve was sampled at 1 Myr intervals (an interval 10 times the 0.1 mA intervals that constitute the fit) as is the biodiversity data (interpolated between adjacent data points only). This approach was also checked against Lomb-Scargle methods (no interpolation) elsewhere (Cornette, 2007; Melott and Bambach, 2011a). We note that our spectral analysis is restricted to periods longer than 25 Myr. Interpolation and sampling effects are only significant for periods much shorter than this (see e.g. Melott and Bambach, 2011a).. All data is resolved to timescales considerably smaller than 25 Myr. We use AutoSignal v.1.7 for power spectra, but cross-spectra were computed with software written by ALM. The online supplementary Appendix contains more background discussion of cross-spectra.

**Results**

In Fig. 1 we show the residuals of marine-$^{87}$Sr/$^{86}$Sr and short-lived genera over most of the Phanerozoic. Their oscillations are visibly anticorrelated; the correlation coefficient is -0.53 and is significant at the level p<0.0000001 against the possibility that noise in uncorrelated series could have given rise to the measured correlation coefficient. A value of -0.44 (p~0.005) results when the same quantity is computed from the count of all marine genera in the independent Paleobiology Database biodiversity (hereinafter PBDB; Alroy et al., 2008; Alroy, 2008).

Fig. 2A shows the power spectrum (computed with AutoSignal 1.7) of the detrended $^{87}$Sr/$^{86}$Sr fluctuations. The highly significant (p<0.001) peak (frequency highlighted on all plots by an arrow marker) appears at 59.3 ± 3 Myr, i.e. within less than one joint standard deviation (the two standard deviations added in quadrature (the total is the square root of the sum of the squares) from the period of the biodiversity fluctuation (Rohde and Muller, 2005; Melott and Bambach (2011a) and references therein). Significance levels include taking account of the length of the record; end effects are removed by zero-padding, detrending, and checking with the alternate method of tapered-cosine windows (Muller and McDonald, 2002; Press et al., 2007). The peak at about 60 Myr even appears in the power spectrum of the non-detrended curve.

In Fig. 2B-D we show the real part of (complex) cross-spectra, since all quantities are nearly in perfect phase or antiphase with one another (see Supplementary Material). Fig. 2B displays the cross-spectrum of $^{87}Sr/^{86}Sr$ with biodiversity. The strong negative peak at ~61 Myr (x-axis value 0.0165 $Myr^{-1}$) results from the anticorrelation seen clearly in Fig.1. The components at the cross-spectral peak are 171 ° out of phase, where 180° would be perfectly so. The coherent fluctuation of $^{87}Sr/^{86}Sr$ ratios and marine biodiversity suggests that geological events with a period of ~60 Myr influence both, and probably share a common cause. Note that the anticorrelation seen in Figures 1 and 2B does not extend to all frequencies. For example, the $^{87}Sr/^{86}Sr$ curve (seen in the supplement, Fig S1) has a long-term trend that is concave downward, over the Phanerozoic. Biodiversity, on the other hand, appears to rise, pause, then rise some more over the same period. This means that the process which drives these variables in opposite directions is specific to a period ~60 Myr.

We next consider the periodicity in the number of sedimentary rock packages mentioned earlier. In Melott and Bambach (2011b) we found that carbonate packages exhibit a much stronger periodicity than do siliciclastic packages, so we regard this analysis of segregated lithologies as more robust than the analysis of Meyers and Peters (2011). The combined standard deviation of the biodiversity result and our carbonate package result, adding in quadrature, is 5 Myr. At 4 Myr difference in period, the two signals are less than one standard deviation apart in frequency. In Melott and Bambach (2011b), Figure 12, the two best-fit sinusoids are plotted together. It can be seen that they coincide closely in the mid-Permian. They are therefore most out of phase at the Cambrian (possibly due to dating ambiguity) and near the present (short-lived genera are undefined after 45 Ma)

In Fig. 2C we show the cross-spectrum between $^{87}Sr/^{86}Sr$ and the number (area and temporal-discreteness weighted) of gap-bound sedimentary packages documented from the United States (Peters, 2006, 2008). The data represent 541 composite stratigraphic columns covering the 48 contiguous United States and Alaska. They define 3221 sedimentary packages that are bounded by unconformities. It is the most complete continent-wide compilation of stratigraphic data available; see especially Peters (2006) for details of his compilation method. Additionally, his p 396-398 discuss the COSUNA (Correlation of Stratigraphic Units of North America) charts used as a source of data. The numbers of these packages oscillate at 58 ± 4 Myr in-phase with biodiversity counts (Melott and Bambach (2011b); see also Meyers and Peters 2011) and is out of phase with $^{87}Sr/^{86}Sr$. For Fig. 2C, we combine carbonate and siliciclastic package data, as both are predominantly marine, but separate analysis of each (Melott and Bambach, 2011b) shows that the former shows a larger amplitude peak at the same period. A peak near 0.01 $Myr^{-1}$ is also present and much more prominent than in B, suggesting that there may be a periodicity around 100 Myr, which does not have a strong effect on biodiversity, in contrast to the one at 58 Myr, which is prominent in both Fig. 2B and 2C.

In Fig. 2D we show the cross-spectrum of eustatic sea level with $^{87}Sr/^{86}Sr$. A peak at ~60 Myr is present, but weak, because the peak for eustasy at this frequency is insignificant, although the one for the $^{87}Sr/^{86}Sr$ spectrum is strong. The eustasy data (from Ogg and Lugowski, 2010) shows only a minor peak in phase with the strong periodicity in marine diversity (Melott and Bambach (2011b)).

In Fig. 3 we show the cross-correlation with $^{87}Sr/^{86}Sr$ of fractions of origination and extinction in the two very different paleontological data sets we have used here. They exhibit very similar patterns, confirming that the correlation is robust and not dependent upon any particular features of either data set. The data sets have different intervals. Since PBDB has longer intervals, its curves are displaced as expected with respect to the Sepkoski data—we assigned originations to the beginnings of intervals, and extinctions to the ends. Therefore this timing is more accurate in Sepkoski data-based curves. Changes in the rate of both origination and extinction contribute to the correlation between biodiversity and $^{87}Sr/^{86}Sr$. A peak means a positive correlation and a trough means a negative correlation. A positive lag means that the isotope ratio precedes the biodiversity. A negative lag means the opposite.

Figure 3A implies that origination is peaking at or just after the isotope peak, which is when biodiversity is a minimum. Plot 3B implies that extinction peaks 10-15 Myr before the isotope ratio peak (when biodiversity is on the way down). These results are fully consistent with our previous analysis (Melott and Bambach, 2011b) of the role of origination and extinction in the biodiversity curves. It does not, however, imply that whatever causes these processes drives both origination and extinction directly. It is possible, for example, that the driver affects extinction rates, and extinction drives origination, as it opens a niche for new species.

Since diversity has a periodicity, it is important to see how the processes that generate diversity and diversity change (origination and extinction) relate to the Sr isotope pattern. The cross-correlation in Fig. 3 documents that both origination and extinction have a role in the periodicity. The relationship of the timing of origination peaks and extinction peaks with the timing of Sr isotope peaks (which negatively correlate with peaks of biodiversity) relate directly to determining the changes in biodiversity with origination peaks at or just after the isotopic peaks (which is when biodiversity is at a minimum) and extinction peaks 10-15 Myr before the isotope peaks (which is during the interval when diversity is decreasing). This pattern also of course connects to the record of number of gap-bound sedimentary packages (which is positively correlated to diversity), too, in that extinction is highest as that pattern is heading for a minimum. Although the last appearances of given genera may simply be timed with loss of record, the fact that these genera never reoccur indicates that the actual extinctions occur during the low point in the record. Originations peak when the record is at its poorest, rather than when it is at its best.

**Discussion**

Variations in values of marine-$^{87}Sr/^{86}Sr$ reflect the interplay of two major, and several minor, influences. The two strongest are continental weathering, which drives marine-$^{87}Sr/^{86}Sr$ to higher values, and hydrothermal circulation at mid-ocean ridges (MOR), which drives it to 6lower values. The hydrothermal flux of Sr from MOR circulation is about half that of continental weathering (Elderfield and Schultz 1996, Table 9). The value of marine-$^{87}Sr/^{86}Sr$ is therefore commonly taken as a proxy for the rate of continental weathering, either through glacial action (Armstrong 1971 et seq.), or increased tectonism (Raymo et al. 1988 et seq.; but see Halverson et al. 2007, Li et al. 2007, for dissenting views). Our finding of a link between $^{87}Sr/^{86}Sr$ and the temporal variation in the number of marine sedimentary packages appears to favor the view that the marine-$^{87}Sr/^{86}Sr$ reflects mostly continental weathering, as changes in relative sea level would cause variation in the area of continent exposed to weathering and the deposition of marine sediments. Indeed, McArthur and Howarth (2001; Fig. 9) noted an overall correspondence between the volume of sedimentation through the Phanerozoic and the major features of the marine-$^{87}Sr/^{86}Sr$ curve: both show Cambrian and Cenozoic maxima, and a Mesozoic minimum.

When continental freeboard is high (and relative sea level is low), epeiric seas are absent or reduced in size and fewer packages of marine sediments can accumulate on continental basement. Because the area of shallow seas can vary over time by an order of magnitude or more (Ogg and Lugowski, 2010), and because of the low correlation with eustasy, we suggest that long-term tectonic processes, which are a likely source of the periodic fluctuation in $^{87}Sr/^{86}Sr$ ratios in sea-water, are also responsible for the periodic waxing and waning in the number of sedimentary packages (Peters, 2008), with low $^{87}Sr/^{86}Sr$ at times of low continental freeboard and high numbers of sedimentary packages and vice versa (Fig. 2C).

We further suggest that times of high continental freeboard and diminished areas of shallow shelf habitat (in which most of the fossil record is deposited) would increase the vulnerability of shallow water marine faunas to stress, especially taxa of limited geographic extent, limited environmental tolerances, or limited species-richness. Since short-lived genera would include most of the genera with limited species numbers, limited geographic and/or environmental distribution, (Jablonski, 1989; Miller, 1997), they would include the genera most likely to be vulnerable to regional uplift. Genera endemic to a region would be the most vulnerable since large scale regional uplift could encompass their entire geographic range. Although some bias in the diversity data may be from temporal fluctuation in the volume of the sedimentary record (more sediments mean more opportunity to find fossils), we have demonstrated in Melott and Bambach (2011b) that the diversity pattern is real and is not an artifact of variable quality of the geologic record. See also Peters (2006, 2008).

The fact that equivalent periodic patterns of fluctuating characterize diversity of short-lived genera, numbers of marine sedimentary packages, and the value of marine- $^{87}Sr/^{86}Sr$ reflects a signal we could call "the pulse of the Earth." The 171° phase angle difference means that $^{87}Sr/^{86}Sr$ is low when biodiversity and the number of sedimentary packages are both high. (The difference between 171° and 180° is well within about one standard deviation of the common timing.) As we will discuss, this may represent the "press" aspect of the press-pulse model (Arens and West, 2008) of mass extinctions (Bambach, 2006) — the idea that they result from periodic background stress combined with catastrophic events (Feulner 2011). Such a combination of forcings can explain the clustering of mass extinction events in the declining diversity phases of the 62 Myr periodicity pattern of fluctuating diversity shown in Melott and Bambach (2011b).

Uplift sufficient to cause significantly enhance erosion of continental rocks is a regional tectonic phenomenon (e.g. mountain building). In addition, glaciation at lower elevations in high-latitude regions can pulverize old cratonic cores and enhance weathering of radiogenic basement (Montañez et al., 2000). Plate-tectonic closure of ocean basins eliminates marine habitats along the colliding coasts (Kearey et al., 2009). Similarly, major regional uplift of continental crust, associated with either crustal thickening or with large-scale intrusions (e.g Karoo sills and dykes; Jourdan et al., 2007), drains shallow seas from the uplifted region. The small signal found in eustatic sea level estimates (Figure 2D) may be due to contamination of the data by effects such as local or regional tectonism (Petersen et al., 2010).

We cannot rule out regional regional tectonism as a cause. The periodicity in the number of sedimentary carbonate packages shown in Melott and Bambach (2011b) and here in Figure 2C is based on North American data. Could this explain a global signal? The $^{87}Sr/^{86}Sr$ signal is global not because the flux of Sr into the oceans is global, but because Sr has a long residence time in seawater relative to the mixing time of the oceans (Hess et al. 1986). Thus large regional fluxes of Sr to the ocean, from the weathering of the Andes for example, will produce a "global" response in the $^{87}Sr/^{86}Sr$ of seawater that is potentially indistinguishable from a more spatially uniform pulse of Sr from intensified climate-driven weathering of the continents. Potential processes that may be regional in extent should not be overlooked.

There is a reported ~60 Myr periodicity in dates on emplacement of large igneous provinces (LIPs) for a few cycles over the last 320 Myr (Prokoph et al., 2004). The multiple effects of increased volcanic activity (added supply of sulfur, carbon dioxide, etc.; Devine et al., 1984), which accompany the genesis of LIPs (Saunders, 2005), could in principle create another stress system (e.g. Sobolev et al. 2011) that might affect the diversity of vulnerable genera, as well as $^{87}Sr/^{86}Sr$ without much affecting eustatic sea level, which would be consistent with the nearly absent signal in Fig 2D. Unfortunately, neither the significance level nor phase of the signal in LIP emplacement times was reported in Prokoph et al. (2004). We have examined the power spectrum of LIP emplacement times in three separate compilations (Prokoph et al., 2004; Arens and West, 2008; and results from the Timescale Creator (Ogg and Lugowski, 2010). We found no significant periodicity at any timescale, and especially no hint of any near 60 Myr (see the Online Appendix, Figure S2).

DeCelles et al. (2009) proposed that periodicity exists in the timing of Cordilleran-type orogens. Their period of ~40–50 Myr was based on visual inspection of geological time series, including Nd isotopes, but the existence of the periodicity is debatable, and requests to release the data for analysis have not been met.

So what might be the reason these phenomena are periodic in behavior? One possible internal earth source of the driver for periodicity is large-scale convective circulation of the mantle. Computational models support the potential for a presence of periodicity in such mantle convection. Periodic upwellings may arise from steady heating (Ribe et al., 2007). Fluid dynamics equations predict recurrence times (with great uncertainty based on uncertain viscosity) of a few 10s of Myr (Davaille and Vatteville, 2005; Ribe et al., 2007; Müller et al., 2008). There is some laboratory experimental support for this sort of behavior (Schaeffer and Manga, 2001).

Other hints of periodic behavior of the earth's interior include evidence for coincident variation of widely separated mantle upwellings with a period of 15 Myr (Mjelde and Faleide, 2009). Small-scale, sublithospheric mantle convection will affect surface movement (Petersen et al., 2010) and so regionally control relative sea level and patterns of erosion and deposition. It has been argued that uplift due to mantle convection provides an alternative to eustatic mechanisms for relative sea level change on timescales up to 100 My (Lovell, 2010). If the mantle convection were periodic, as claimed based on the computational models, this would explain periodic uplift, changes in continental freeboard and sedimentation, with the observed absence of a significant periodic signal in eustatic sea level (Melott and Bambach, 2011b). Since the $^{87}Sr/^{86}Sr$ signal is a global one, presumably a globally synchronized periodicity in mantle convection would be required—an idea which finds some support in the observations of Mjelde and Faleide (2009) that there has been a synchronization between Iceland and Hawaii. Gyüre and Jánosi (2009) show that whole-mantle convection (rather than isolated up- and downwellings) with a period of a few tens of millions of years is the expected behavior.

Hints of periodic behavior of production of oceanic sea-floor and, therefore, eustatic variation from changes in oceanic basin volume have been described (Becker et al., 2009), but we have been unable to find the claimed ~60 Myr period in that data. Other evaluations of sea-floor spreading rate for the past 140 Ma (e.g. Seton et al., 2009) differ considerably, so no firm conclusions can be reached about any periodicity in sea-floor spreading rates, although this is another system that likely plays a role in the phenomena we discuss.

Climate is another possible common driver of bidoversity, $^{87}Sr/^{86}Sr$, and deposition of sedimentary packages, since it can affect erosion rates (e.g. Li et al., 2007). There is intense debate over the relative contributions of tectonics and other erosion to $^{87}Sr/^{86}Sr$ change (Molnar and England, 1990; Dowdeswell et al., 2010; Egholm et al., 2009), but effects of increased erosion and subsequent uplift due to unloading by retreat of ice-caps will affect both $^{87}Sr/^{86}Sr$ and continental freeboard. This driver is potentially interesting because there are possible climate drivers that have a built in periodicity at ~62 Myr. Medvedev and Melott (2007) have shown that the oscillation of the Solar System normal to the galactic disc with a period of ~62 Myr should result in increased exposure to high-energy cosmic rays when the Earth is north of the galactic disc (north taken as the maximum of Galactic latitude, in the general direction of the Virgo Cluster). Atmospheric ionization caused by such cosmic rays has been proposed as a causal agent in changing terrestrial cloud cover (Melott et al. 2010). Climate change particularly at high altitude may changee erosion rates and the $^{87}Sr$ flux into the oceans. Calculations indicate that radiation propagating to the ground may be sufficient to significantly increase the radiation load from muons on organisms on the surface and the top ~1km of ocean (Atri and Melott, 2011a, 2011b). This is a speculative but physically reasonable mechanism to lower biodiversity that gives the right period and phase, based on solutions of the motion of the solar system in the Galaxy (Gies and Helsel, 2005).

The data and our analyses document that a highly regular periodic process affects both the evolution of life in the oceans and some geochemical and sedimentary systems. While we have hints that can connect tectonic systems (or possibly astronomical ones) to these Earth surface phenomena, we are left with a theoretical challenge: we have no exclusive theory for periodic processes internal to the Earth. What is the driving mechanism and why is it periodic in effect?

**Acknowledgments**

We thank J. Alroy, P. Flemings, J. Ogg, and S. Peters for providing some of the data analyzed here. T. Hallam made useful comments. An anonymous referee and Michael Foote provided extremely helpful feedback in review. We are grateful to the American Astronomical Society for the sponsorship of the 2007 Honolulu multidisciplinary Splinter Meeting at which discussions leading to this project took place.

Sobolev, S. V.; Sobolev, A. V.; Kuzmin, D. V.; Krivolutskaya, N. A.; Petrunin, A. G.; Arndt, N. T.; Radko, V. A.; and Vasiliev, Y. R. 2011. Linking mantel plumes, large igneous provinces, and environmental catastrophes. Nature 477: 312-316.

**FIGURE CAPTIONS**

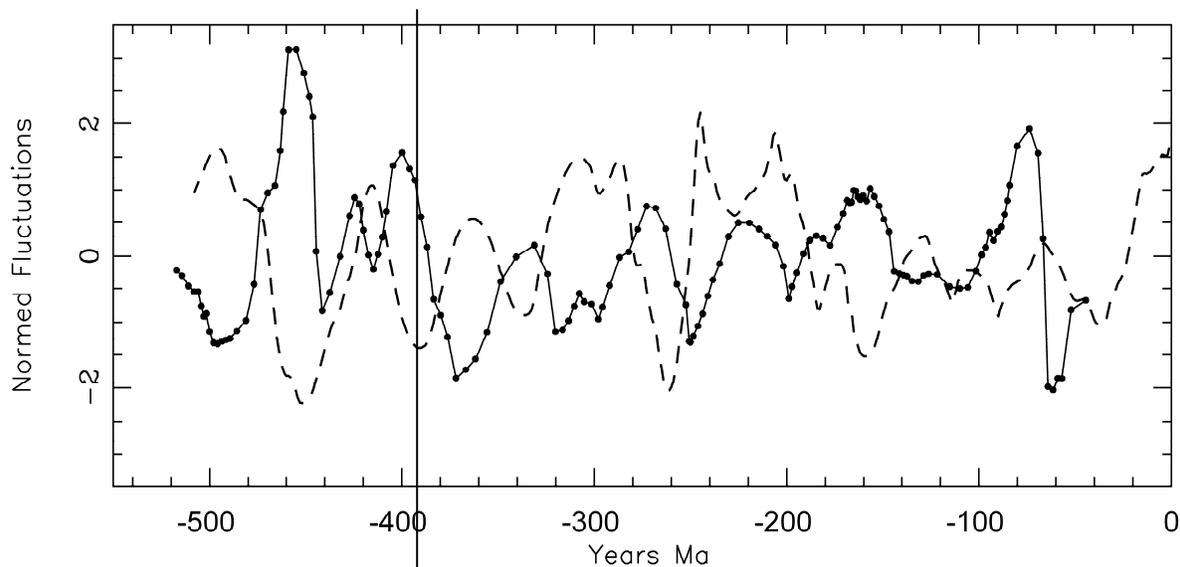

**Fig. 1** Diversity of genera which lasted less than 45 Myr, and marine-$^{87}Sr/^{86}Sr$ values. Each curve has been detrended and divided by its standard deviation to put them on a common y axis range. The dashed line is the best-fit to the $^{87}Sr/^{86}Sr$ ratio, and the solid line with points is the number of genera of short-lived fauna (see text).

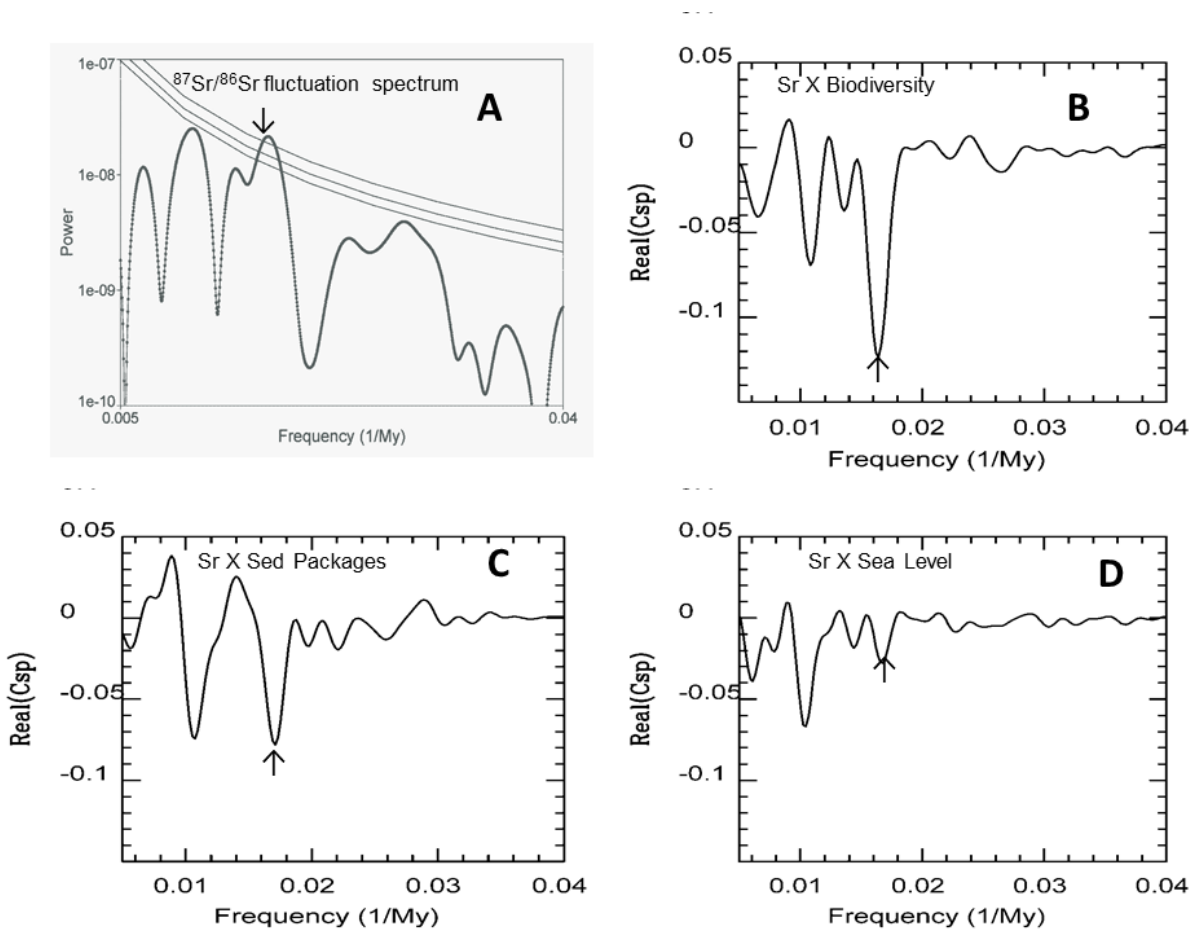

**Fig. 2:** All four plots have semilog axes. (A) Power spectrum of the fluctuations in detrended marine $^{87}Sr/^{86}Sr$ values. The x-axis label units are frequency, the inverse of the cycle time, as is customary with spectra. The parallel lines indicate significance levels of p = 0.05, 0.01, and 0.001 respectively. The peak at 59.3 ± 3 Myr is significant at p < 0.001, and is responsible for 28% of the variance in the detrended data. (B) The real part of the cross spectrum between detrended biodiversity of short-lived genera (62 ± 3 Myr, see text) and $^{87}Sr/^{86}Sr$, showing a strongly negative peak at ~61 Myr—which implies that biodiversity and the isotopic ratio *oscillate out of phase* on these time scales. (C) The cross-spectrum between the detrended $^{87}Sr/^{86}Sr$ and the (area weighted) number of marine gap bound sedimentary packages (period 58 ± 4 Myr). Note the existence of strong negative peaks at 58 and 92 Myr, which are, again, *out of phase* between the two signals. More sedimentary packages being deposited at a given time means lower $^{87}Sr/^{86}Sr$, for variations on these timescales. (D) Cross-spectrum of the $^{87}Sr/^{86}Sr$ isotopic ratios with sea level (Ogg and Lugowski, 2010). The power spectrum of the sea level data alone has a broad peak at a period of 58-67 Myr, but it is far too low to be considered significant (Melott and Bambach (2011b), Figure 11). Some insignificant inverted phase oscillation occurs at frequency of 60 Myr, which may reflect contamination of the eustatic signal by correlated tectonic uplift effects. Stand-alone power spectra for biodiversity are shown in Melott and Bambach (2011a), and those for sedimentary packages (C) and sea level (D) are shown in Melott and Bambach (2011b), Figures 10 and 11 and are not repeated here.

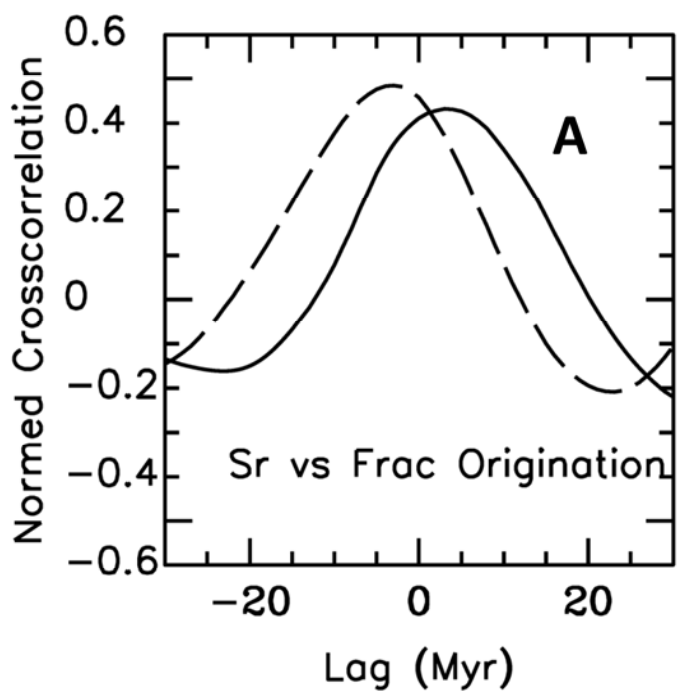

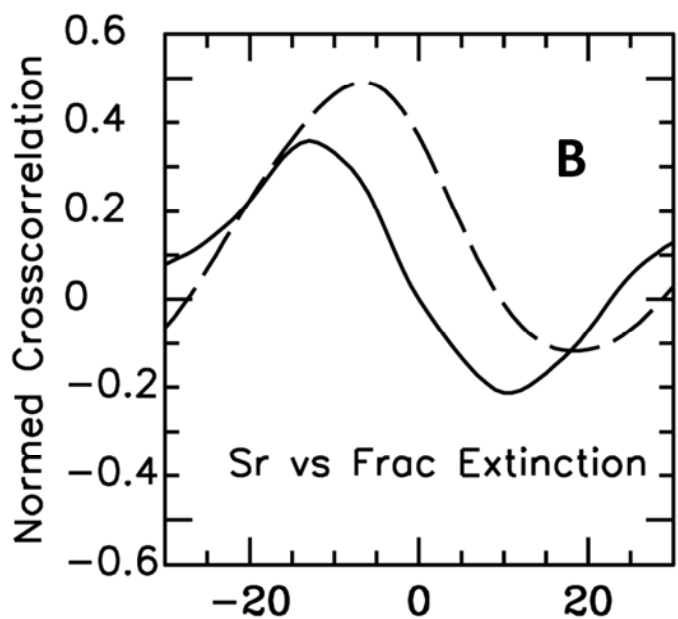

Fig 3. The cross-correlation as a function of lag in Myr between detrended and normalized variables $^{87}Sr/^{86}Sr$ and fractional origination and extinction (origination or extinction divided by the total number of genera extant at a given stage) in our two paleontological datasets. Positive lag is defined as having the isotopic ratio change preceed the biodiversity change. The solid lines represent the fraction of origination and extinction of short-lived genera in the Sepkoski (2002)

data, as plotted in Fig. 1 from Melott and Bambach (2011a). The dashed lines represent similar analysis of the variables λ and μ which describe extinction and origination rates in the PBDB as specifically defined elsewhere (Alroy et al., 2008; Alroy, 2008). The upper panel (A) is origination, the lower panel (B) extinction. Similar patterns are exhibited between the two independent and different data sets, confirming the robustness of the relationship with $^{87}Sr/^{86}Sr$. The common pattern confirms that the interaction of origination and extinction is needed to establish large swings in biodiversity, and that both are correlated with $^{87}Sr/^{86}Sr$ fluctuations. The curves are shown over a lag interval of ± 30 Myr, and are nearly periodic over the plotted domain, further confirming the centrality of the ~60 Myr periodicity in marine $^{87}Sr/^{86}Sr$. Extinctions were assigned to the ends of intervals in which the data was binned, and origination was assigned to their beginnings. This results in a time displacement of about 7 Myr between the curves based on the two sets of data, as the PBDB data has a much longer mean interval.

# Appendix to

Melott, Bambach, Petersen, and McArthur, "**A ~60 Myr periodicity is common to marine-$^{87}$Sr/$^{86}$Sr, fossil biodiversity, and large-scale sedimentation: what does the periodicity reflect?**"

I Methods

In the main text, we mention that the quantities considered here were detrended before proceeding with time series analysis. For a variety of reasons, this is customary in time series analysis. Not the least of these is moving the focus from long-term secular change to fluctuations about that change. We show here the process for the strontium isotope ratios. In Figure S1, we show the LOWESS fit and its detrending line.

The detrending line is a cubic regression, over the time period for which substantive fossil diversity data exist. The regression we present here is the best fit per degree of freedom within polynomials and transcendental functions. The fluctuations shown in Figure 1 are the residuals of the data around the best-fit cubic, divided by their standard deviation in order to put them on an equal footing with the biodiversity data shown in the same plot.

Time series analysis is performed by Fourier Transform, using the widely distributed IMSL software. This decomposes any function into a sum over sinusoidal waves of different frequencies. The question of interest is whether any of these stands out at significant high amplitude, suggesting a periodicity. The power spectrum shown in Fig. 2A is the diagnostic of this.

The cross-spectral coefficient C is a complex number generalization of the power spectrum for two time series. These complex numbers can be represented as $re^{i\theta}$, where r is the product of amplitudes with which the same frequency is represented in the two series, and $\theta$ is the phase angle between the two oscillations. If $\theta$ is zero, they are in phase and C is real and positive. Imaginary numbers result from 90° phase angles, negative real numbers from 180° phase angles, and of course in general complex numbers from other angles.

We are further interested in the relationship between two time series. more specifically whether they have a strong joint variation at a particular frequency. For this we use the cross-spectrum. This is done by analysis of the cross-spectrum (Press et al., 2007) of the two detrended series. It is most easily described using the complex exponential representation of Fourier Series. This is a generalization of the power spectrum of a single time series, which is essentially $A_i{}^*A_i$, where * denotes complex conjugation and $A_i$ denotes elements of a series of complex Fourier coefficients as a function of frequency. The power spectrum is therefore real-valued. The cross-spectrum involves the Fourier coefficients of two different series: $B_i{}^*C_i$, and it is complex. The amplitude of this complex number is a measure of the extent to which a given frequency is present in both series; its phase is a measure of the extent to which the components of the two series are in phase, i.e. whether the timing of the peaks and troughs coincide. To the extent that the signals are anchored in objective reality, we expect phase agreement between components which came from different data sets. If they are in perfect phase, the cross-spectral coefficient is real and positive near the peak. Out of phase signals with the same frequency will not be positive real numbers—they are complex numbers, and may have a negative real part or have a substantial imaginary component. Thus the test for a positive peak in the real part of the cross-spectrum is a stringent test that requires not only that both data sets have the same frequency at strength but also that the signals have some component in phase—the peaks and troughs coincide. Similarly, a negative real peak, as seen several times in this work, is indicative of two cycles out of phase. We can also examine the imaginary part and real part together, and determine the phase lag between the two time series at a given frequency.

## II. Periodicity test on LIP Emplacement Times

In the main text, we mentioned that we found no evidence of a 60 Myr periodicity in the emplacement times of Large Igneous Provinces. We examined three compilations (1) Data meeting the criteria for sample AB10 in Propkoph et al. (2004), compiled from Table 1. These are events with a strong probability of being associated with an arriving mantle plume head, and a (±2σ) dating uncertainty of less than 10 Myr. (2) Data taken from Appendix 3 of Arens and West (2008), comprising continental flood basalts. (3) Data downloaded from Timescale Creator (Ogg and Lugowski 2010) and found no significant spectral peaks. We present here in Figure S2 these spectra as an example typical of the results.

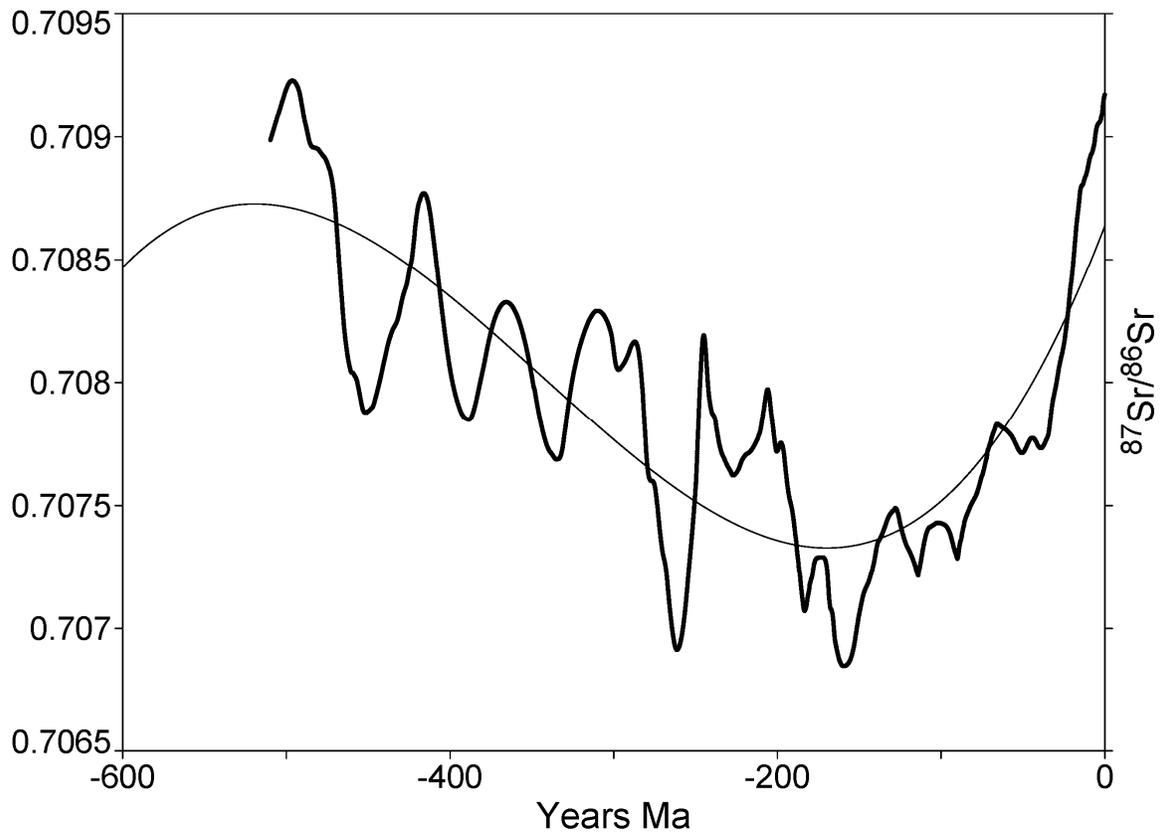

Fig. S1: The strontium isotope ratios specified in the text are plotted as the heavy line. The best-fit cubic (determined by least squares fitting) is the light line, and the residuals of the data around the best-fit cubic are the quantities plotted in Fig. 1 (after dividing by the standard deviation).

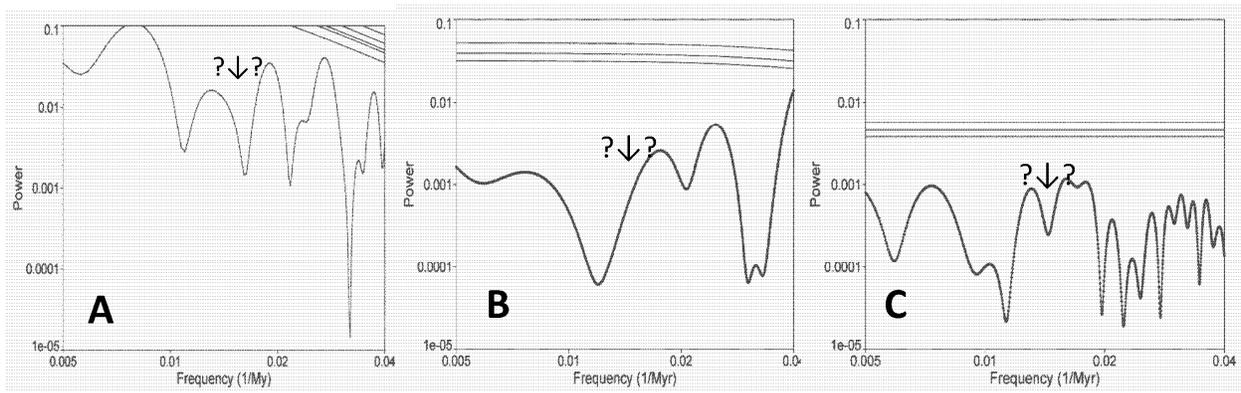

Fig. S2: (A) The solid line represents the power spectrum of temporal fluctuations in the rate of LIP emplacements in the AB10 data of Prokoph et al. (2004), Table 1. To be significant, the values would need to intersect the lines of significance (5%, 1%, 0.1%) visible in the upper right corner of the plotted area.  Furthermore, there is no significant power at a frequency of 0.016, marked by an arrow, which would correspond to the ~60 Myr periodicity of interest. (B) Same as A, but for the Arens and West (2008) LIP emplacement table. (C) Same as A, but for a table downloaded from the Timescale Creator, (Ogg and Lugowski, 2010). Differences in the slope of significance standards result from fits to the overall spectral shape, outside the bounds of the plots.